\documentclass[preprintnumbers,prd,showpacs,floatfix,superscriptaddress,nofootinbib,twocolumn]{revtex4-1}
\usepackage{graphicx,epsfig}
\usepackage{amsmath}
\usepackage {amssymb}
\usepackage {longtable}
\usepackage{multirow}
\usepackage{dcolumn}
\usepackage{bm}
\usepackage{amsfonts}
\usepackage{subfigure}
\usepackage{color}
\usepackage{relsize}

\newcommand{\be}{\begin{equation}}
\newcommand{\ee}{\end{equation}}
\newcommand{\ben}{\begin{eqnarray}}
\newcommand{\een}{\end{eqnarray}}
\newcommand{\bes}{\begin{subequations}}
\newcommand{\ees}{\end{subequations}}
\newcommand{\bens}{\begin{subeqnarray}}
\newcommand{\eens}{\end{subeqnarray}}

\def\tanh{\text{tanh}}

\def\Ci{\text{Ci}}

\begin{document}

\title{Novel modified gravity braneworld configurations with a Lagrange multiplier}

\author{Dionisio Bazeia}
\email{dbazeia@gmail.com}
\affiliation{Departamento de F\'isica, Universidade Federal da Para\'iba, 58051-970 Jo\~ao Pessoa, PB, Brazil}
\author{D. A. Ferreira}
\email{ferreira.douglasdaf@gmail.com}
\affiliation{Unidade Acad\^emica de F\'isica,
Universidade Federal de Campina Grande,
58429-900 Campina Grande, PB, Brazil}
\author{Francisco S. N. Lobo}
\email{fslobo@fc.ul.pt}
\affiliation{Departamenteo de Física, Faculdade de Ci\^encias da Universidade de Lisboa, \\
Instituto de Astrofísica e Ci\^encias do Espa\c{c}o, Faculdade de Ci\^encias da Universidade de Lisboa, Edif\'icio C8, Campo Grande, P-1749-016, Lisbon, Portugal}
\author{Jo\~{a}o Lu\'{i}s Rosa}
\email{joaoluis92@gmail.com}
\affiliation{Institute of Physics, University of Tartu, W. Ostwaldi 1, 50411 Tartu, Estonia}

\date{\today}

\begin{abstract}

In this work we deal with thick brane solutions in the warped five-dimensional braneworld scenario with a single extra spatial dimension of infinite extent, for a class of modified theories of gravity with a Lagrange multiplier. We first present the action, describe the gravity and field equations, outline a strategy to find explicit solutions and explore the stability of the gravitational sector. The investigation deals mainly with the construction of a first order framework capable of using a single scalar field to simulate warp functions that appear in two-field models. In particular, we find specific symmetric and asymmetric brane configurations that engender asymptotic profiles with symmetric and asymmetric five-dimensional anti-de Sitter geometries. Thus, including a Lagrange multiplier unveils an alternative approach to induce brane structure using a single scalar field, tracing out new avenues of research in braneworld scenarios, naturally leading to interesting results for the localization of matter fields in the brane.

\end{abstract}

\maketitle

\section{Introduction}

A higher-dimensional mechanism for solving the Hierarchy Problem was proposed within the Randall-Sundrum braneworld scenario that describes the four-dimensional observable universe as a localized brane embedded in a higher dimensional spacetime, denoted the {\it bulk} \cite{Randall:1999ee,Randall:1999vf}. In braneworld phenomenology, the Standard Model fields are contained on the brane and gravity propagates freely in the bulk. Shortly after the Randall-Sundrum scenario, a mechanism for stabilizing the size of the extra dimension was proposed, where the potential for the modulus field, that sets the size of the fifth dimension, is generated by a bulk scalar with quartic interactions localized on the two three-branes  \cite{Goldberger:1999uk}. In fact, it was shown that the minimum of this potential yields a compactification scale that solves the hierarchy problem without fine tuning of parameters.

However, solutions to the classical equations for scalar fields in five dimensions can also generate topological structures even in the absence of gravity, and thus induce localized brane scenarios \cite{DeWolfe:1999cp}. The method is inspired by gauged supergravity and involves only first order differential equations, with several interesting applications, namely, taking into account a full nonlinear treatment for the stabilization mechanism of interbrane spacing and the construction of smooth domain wall solutions. In addition, the study of the localization of four-dimensional gravity in a simple version of a thick domain wall in ${\rm AdS_5} $ space was considered in \cite{DeWolfe:1999cp,Gremm:1999pj,Csaki:2000fc}.

Generalizations of the Randall-Sundrum model were also explored with and without scalar fields, that include three-branes in higher dimensional spaces which are not necessarily anti-de Sitter far from the branes, intersecting brane configurations and configurations involving negative tension branes \cite{Csaki:2000fc}. Furthermore, it was also found that three-brane metrics in five dimensions can arise from a single scalar field source. Scalar fields have also been studied to build thick brane structures in several other contexts \cite{Brito:2001hd,Kehagias:2000au,Campos:2001pr,Bazeia:2003aw}. For example, they have been explored in supergravity theory \cite{Brito:2001hd} and as five-dimensional bounce solutions to the Einstein equations coupled to a scalar field, associated with ${\rm AdS_5}$ spaces with smooth warp functions \cite{Kehagias:2000au}, where the dilaton, as a bulk scalar leads, through its coupling, to localized gauge boson fields. Also, the effects of a generic bulk first-order phase transition on thick Minkowski branes in warped geometries have also been investigated \cite{Campos:2001pr}. It is interesting that models described by scalar potentials that drive the system to support thick brane linearly stable solutions, that engender internal structure, have also been explored \cite{Bazeia:2003aw}. 

The subject started in \cite{Randall:1999ee,Randall:1999vf} gave rise to a braneworld scenario in which the brane is thin, but it was further investigated in several works, in particular, in Refs.  \cite{Goldberger:1999uk,DeWolfe:1999cp,Gremm:1999pj,Csaki:2000fc,Brito:2001hd,Kehagias:2000au,Campos:2001pr,Bazeia:2003aw}, in which the presence of source scalar fields has led to the thick brane scenario. This inspired several new possibilities, some of them with the usage of two scalar fields coupled with gravity. In particular, a Bloch-type domain wall has been considered as a brane candidate, where the parameter which controls the way the two scalar fields interact induces the appearance of a thick brane which engenders internal structure, driving the energy density to localize inside the brane in a very specific way \cite{Bazeia:2004dh}. Since then, several other studies on Bloch branes have appeared with distinct motivations; see, e.g., Refs.  \cite{deSouzaDutra:2008gm,Dutra:2013jea,Bazeia:2016uhr,Almeida:2018bzx,Brito:2019bbb,Almeida:2009jc,Castro:2010au,Cruz:2013zka,Xie:2015dva,Cruz:2012kd,Xie:2013rka,Zhao:2014gka} and references therein. For instance, extended versions of the Bloch brane concept were proposed in \cite{deSouzaDutra:2008gm,Dutra:2013jea,Bazeia:2016uhr,Xie:2015dva,Almeida:2018bzx,Brito:2019bbb}, and the localization of distinct types of matter and fields have been investigated  \cite{Almeida:2009jc,Castro:2010au,Cruz:2013zka,Xie:2013rka,Cruz:2012kd,Zhao:2014gka,Xie:2015dva}. Furthermore, in recent work \cite{Bazeia:2020qxr}, another mechanism to control the two scalar fields was considered, in which one includes a function of one field that modifies the kinematic evolution of the other field. This was inspired by Ref. \cite{Bazeia:2019vld} in flat spacetime, and in the braneworld context described in \cite{Bazeia:2020qxr}, it was shown that the brane may engender interesting new characteristics, in particular the presence of multiple internal structures.

 Another issue of current interest concerns the asymmetric braneworld scenario, in which the brane is immersed in a five-dimensional spacetime with a single extra dimension of infinite extent and asymptotically connects spacetimes with distinct cosmological constants. The asymmetric profile of the brane has been studied with distinct motivations, for instance, to investigate the cosmic acceleration in \cite{Padilla:2004tp,Padilla:2004mc}, the hierarchy problem \cite{Dutra:2013jea,Dutra:2014xla}, and new mechanisms for gravity localization \cite{Gregory:2000jc,Dvali:2000rv,Csaki:2000pp}. Thick asymmetric braneworld scenarios generated by adding a constant to the superpotential associated with the scalar field, have also been considered  \cite{Bazeia:2013usa}, and generalized versions of the Randall-Sundrum model \cite{Randall:1999vf} with different cosmological constants on each side of a brane has been discussed \cite{Ahmed:2013mea}. The stability of asymmetric thick brane solutions in the warped five dimensional braneworld scenarios with a single extra spatial dimension of infinite extent have been explored, and it was shown that the solutions are gravitationally stable against small perturbations of the metric \cite{Bazeia:2019avw}. 
 
 There is another line of investigation of current interest, which refers to the study of thick branes within the context of modified gravity. This has been considered in several different contexts in Refs. \cite{Afonso:2007gc,Dzhunushaliev:2009dt,Zhong:2010ae,Liu:2012rc,Bazeia:2013uva,Menezes:2014bta,Bazeia:2014poa,Fu:2016szo,Rosa:2020uli,Bazeia:2015owa,Fu:2016rgr,Zhong:2017uhn,Mazani:2020abe,Bazeia:2020zut} and in references therein. As is well known, modified theories of gravity in the presence of a Lagrange multiplier may impose important restrictions in the derivative of the other field \cite{A,B,C}. In Ref. \cite{A}, for instance, the authors proposed a model, in which one may unify the description of dark matter and dark energy in the form of a dusty dark energy model. Furthermore, in \cite{B} the authors showed that dark energy cosmology of different types may be reconstructed in such models, and in \cite{C} the investigation suggests that the presence of Lagrange multiplier may be used to generate cold dark matter from quintessence. As we can see, these models are of current interest since they may contribute to produce a unified description of dark matter and dark energy and more, an interesting scenario arises, in particular, for the construction of braneworld solutions in modified gravity with Lagrange multipliers \cite{Zhong:2017uhn,Bazeia:2020zut}. The general aspects of the model were studied and a useful first-order formalism was presented in \cite{Bazeia:2020zut} to find analytic solutions of the equations of motion. Furthermore, explicit models were analyzed and the linear stability of the metric investigated. 

In this work, we build on the latter work \cite{Bazeia:2020zut}, and for this we organize the paper in the following manner: In Sec. \ref{sec2}, the action and gravitational field equations are presented, a first order framework strategy to find explicit solutions is outlined and the stability of the gravitational sector is explored. 
The novelty here is the obtainment of first order differential equations in the presence of a single scalar field, which allows that we discuss, in Sec. III, some specific models. The results show that we are able to use a single field to simulate models described by two scalar fields, and find specific solutions, such as  brane configurations that asymptotically connect two ${\rm AdS_5}$ geometries with different cosmological constants. Finally, in Sec. \ref{sec:Conclusion} we discuss our results and comment on some new lines of investigations of current interest to braneworld with a Lagrange multiplier.

\section{General formalism} \label{sec2}

\subsection{Spacetime metric}

Throughout this work, we consider that the background geometry has a four-dimensional Poincar\'e invariance and a single extra dimension of infinite extent. Thus, the five-dimensional spacetime metric is given by the general {\it ansatz}
\be \label{warped}
ds_{5}^{2}=e^{2A(y)}\eta_{\mu\nu} dx^{\mu}dx^{\nu}-dy^{2}\,,
\ee
where the coordinate $y$ denotes the extra dimension, $A(y)$ is the warp function ($e^{2A}$ is denoted the warp factor used throughout this work), and $\eta_{\mu\nu}=\text{diag}(1,-1,-1,-1)$ describes the four-dimensional Minkowski metric with
$\mu,\nu=0,1,2,3$. Thus, the five-dimensional spacetime metric is given by $g_{ab}={\rm diag}\left( e^{2A}, -e^{2A}, -e^{2A}, -e^{2A}, -1 \right)$, where the latin index runs from $a,b=0,1,2,3,4$.

\subsection{Action and gravitational field equations}

Here, we consider thick brane structures generated by a single scalar field described by the action featuring a Lagrange multiplier, given by
\be \label{action}
S=\!\int \! d^{5}x\sqrt{|g|}\left[-\frac{R}{4}+I\left(\frac{1}{2}\partial_{a}\phi\partial^{a}\phi+\!V(\phi)\!\right)\! - U(\phi)\!\right]\!,
\ee
where $g$ is the determinant of the background metric in Eq. (\ref{warped}), $R$ is the five-dimensional curvature scalar, the two potentials $V(\phi)$ and $U(\phi)$ are functions of the scalar field $\phi$, and $I$ represents the Lagrange multiplier. For simplicity, we consider that the scalar field is static and depends only on the extra spatial dimension $y$. Note that, as we will later verify, the potentials will essentially model the brane and will be involved in the specific scalar field solutions. 
 
In order to deduce the equations of motion, we now consider the variation of the action given in Eq. \eqref{action} with respect to the scalar field $\phi$, the background metric $g_{ab}$, and the Lagrange multiplier $I$, which provide the following field equations:
\bes\label{fe}\ben\label{eom}
I\left(\square\phi-\dfrac{dV}{d\phi}\right)+\partial_{a}I\partial^{a}\phi+\dfrac{dU}{d\phi}=0,\\[3pt] \label{einstein}
G_{ab}-2T_{ab}=0,\\[3pt] \label{constrain}
\frac{1}{2}g^{ab}\partial_{a}\phi\partial_{b}\phi+V(\phi)=0\,,
\een\ees
respectively, where $\square=g^{ab}\nabla_{a}\nabla_{b}$. 
Note that Eq. \eqref{constrain} imposes a constraint on the scalar field $\phi$ to be specified along with the potential $V(\phi)$, which is used to simplify the energy-momentum tensor in the form
\be\label{em}
T_{ab}=I\partial_{a}\phi\partial_{b}\phi+g_{ab}U(\phi).
\ee
Thus, the Lagrange multiplier can be interpreted as a field source, effectively modifying the scalar field dynamics.

Now, using the spacetime metric from Eq. (\ref{warped}), Eqs. \eqref{eom} and \eqref{einstein} yield the following
\bes\label{systeq1}\ben \label{fieldeq}
I\left(\phi''+4A'\phi'+\frac{dV}{d\phi}\right)+I'\phi'-\frac{dU}{d\phi}=0,\\[3pt]
\label{E3}
3A'^{2}+U(\phi)-I\phi'^{2}=0,\\[3pt]
\label{E2}
A''+\frac{2}{3}I\phi'^{2}=0,
\een\ees
respectively, and the constraint in Eq. \eqref{constrain} takes the form
\begin{equation}
\label{constrain2}
\phi'^{2}=2 V(\phi),
\end{equation}
where the prime denotes a derivative with respect to $y$. 

We now have a system of four differential equations, that are not entirely independent
as Eq. \eqref{fieldeq} can be obtained by taking into account Eqs. \eqref{E2} and \eqref{constrain2}, and deriving Eq. \eqref{E3} with respect to $y$.

\subsection{Strategy to find solutions: Auxiliary functions} 
 
In order to find solutions to the above configuration, we now consider the first order formalism that was originally proposed in Ref.~\cite{Bazeia:2020zut}. To this effect, assume that the potentials $V(\phi)$ and $ U(\phi)$ have the specific forms
\be \label{potV}
V(\phi)=\frac{1}{2}\left(\dfrac{d\omega}{d\phi}\right)^{2},
\ee
and 
\be\label{potU}
U(\phi)=\dfrac{d\omega}{d\phi}\dfrac{dW}{d\phi}-\frac{4}{3}\,W^{2},
\ee
where $\omega=\omega(\phi)$ and $W=W(\phi)$ are smooth auxiliary functions of the scalar field. In this case, we obtain the following first order differential equations
\be\label{phifo}
\phi'=\dfrac{d\omega}{d\phi}\,,
\ee
and
\be\label{warpfo}
A^\prime=-\frac23 W\,,
\ee
which are compatible with Eqs. \eqref{systeq1} and \eqref{constrain2}. For this case, the functions $\omega(\phi)$ and $W(\phi)$ are related by 
\be\label{LM} 
I\frac{d\omega}{d\phi}=\frac{dW}{d\phi}.
\ee 
Note that the auxiliary functions $W(\phi)$ and $\omega(\phi)$ are extremely useful as, in addition to inducing the expression for the warp factor, they dictate how the scalar field self-interact and affect the dynamics of the scalar field. 

For the specific case of $W(\phi)=\omega(\phi)$, Eq. \eqref{LM} imposes that the Lagrange multiplier to be a unit, and consequently, the action in Eq. \eqref{action} corresponds to the standard model with the potential 
\be
U_S(\phi)=U(\phi)-V(\phi),
\ee
and Eqs. \eqref{potV} and \eqref{potU} lead to 
\be  
U_S(\phi)=\frac12\left(\frac{dW}{d\phi}\right)^2-\frac43\; W^2,
\ee
which is the form required for the potential of the standard gravity braneworld model, capable of generating thick branes in the presence of the source scalar field, governed by the first order equations, Eqs. \eqref{phifo} and \eqref{warpfo}. 

Using the auxiliary functions, it is also straightforward to obtain the energy density of the system from the energy-momentum tensor in Eq. \eqref{em}, which yields
\be\label{d.energy}
T_{00}=\rho(y)=e^{2A}U,
\ee
and under the use of the first order equations, can be expressed as the following total derivative 
\be 
\rho(y)=\frac{d}{dy}(We^{2A}).
\ee

\subsection{Stability of the gravitational sector}

An important issue that we now investigate is related to the stability of the gravitational sector of the models. To this effect, consider the redefinition $dy=e^{A(z)}dz$ in Eq.~\eqref{warped}, so that the background geometry is conformally flat, i.e., $ \tilde{g}_{ab}=e^{2A(z)}\eta_{ab}$. Then assume small fluctuations around the metric in the form $\eta_{ab}\to\eta_{ab}+h_{ab}(x,z)$, with $h_{44}=0$, so that the perturbed metric is given by
\be \label{mp}
ds^{2}=e^{2A(z)}(\eta_{ab}+h_{ab})dx^{a}dx^{b}\,.
\ee
Using the transverse traceless gauge, i.e., $\partial_\mu h^{\mu\nu}=0$ and $h_\mu^\mu=0$, the linearized Einstein equation yields
\be \label{2.3}
\partial_{c}\partial^{c}h_{\mu\nu}-3{\dot A}{\dot h}_{\mu\nu}=0,
\ee
where the overdot represents a derivative with respect to the new coordinate $z$. In order to understand how the above equation relates to the stability of the system, consider $h_{\mu\nu}=e^{ipx}e^{-3A/2}H_{\mu\nu}$, which leads to the following Schr\"odinger-like equation
\be \label{2.4}
\left[-\partial^{2}_{z}+u(z)\right] H_{\mu\nu}=p^{2}H_{\mu\nu}\,.
\ee
The quantity $u(z)$ is the stability potential, which is given by
\be \label{stability1}
u(z)=\frac{3}{2}{\ddot A}+\frac{9}{4}{\dot A}^{2}\,.
\ee
Now, Eq. \eqref{2.4} can be factorized as
\be
Q^\dag Q H_{\mu\nu}=p^2 H_{\mu\nu}\,,
\ee
with the operators $Q$ and $Q^\dag$ given by
\be  
Q=-\partial_z+\frac32 {\dot A}\,\qquad \text{and} \qquad Q^\dag=\partial_z+\frac32 {\dot A}\,.
\ee
This factorization implies that $p^{2}\geq0$, i.e., the stability equation can only support states with non-negative eigenvalues. In this sense, the braneworld system is stable under small perturbations of the metric.

The graviton zero mode ($p^2=0$) is obtained from $Q H_{\mu\nu}^{(0)}=0$, which results in
\be
H_{\mu\nu}^{(0)}=N_{\mu \nu} e^{3A(z)/2}\,,
\ee
where $ N_{\mu \nu} $ is a normalization factor. Localized four-dimensional gravity then requires the zero mode to be normalizable, i.e., 
\be\label{normcond}
\int_{-\infty}^{\infty} dz\, e^{3A(z)}=\int_{-\infty}^{\infty} dy\, e^{2A(y)}<\infty\,.
\ee
This condition implies that the warp factor must go to $0$ as $ y\rightarrow \pm \infty $.

\section{Some Specific Models} \label{sec3}

In this section, we discuss some models with a single scalar field, which are able to simulate models described by two scalar fields that have been studied in the literature, as we illustrate below. 

\subsection{First model} \label{sub1}

Let us consider the following pair of functions
\bes\label{wW1}\ben\label{w1}
\omega(\phi)&=&2r\left(\phi-\frac{\phi^{3}}{3}\right),\\[3pt] \label{W1}
W(\phi)&=&2r\phi-2\phi^{3}\left(r-\frac{1}{3}\right)+c\,,
\een\ees
where $r$ and $c$ are real constants. Note that, if $r =1/2$ and $c=0$, the two functions above are the same and we get the standard thick brane scenario described by the well-known $\phi^{4}$ model \cite{DeWolfe:1999cp}.

\begin{figure}[htb!]
\centering
\subfigure[~Scalar field solution $\phi(y)$] {\label{fig1a}%
\includegraphics[width=7.2cm]{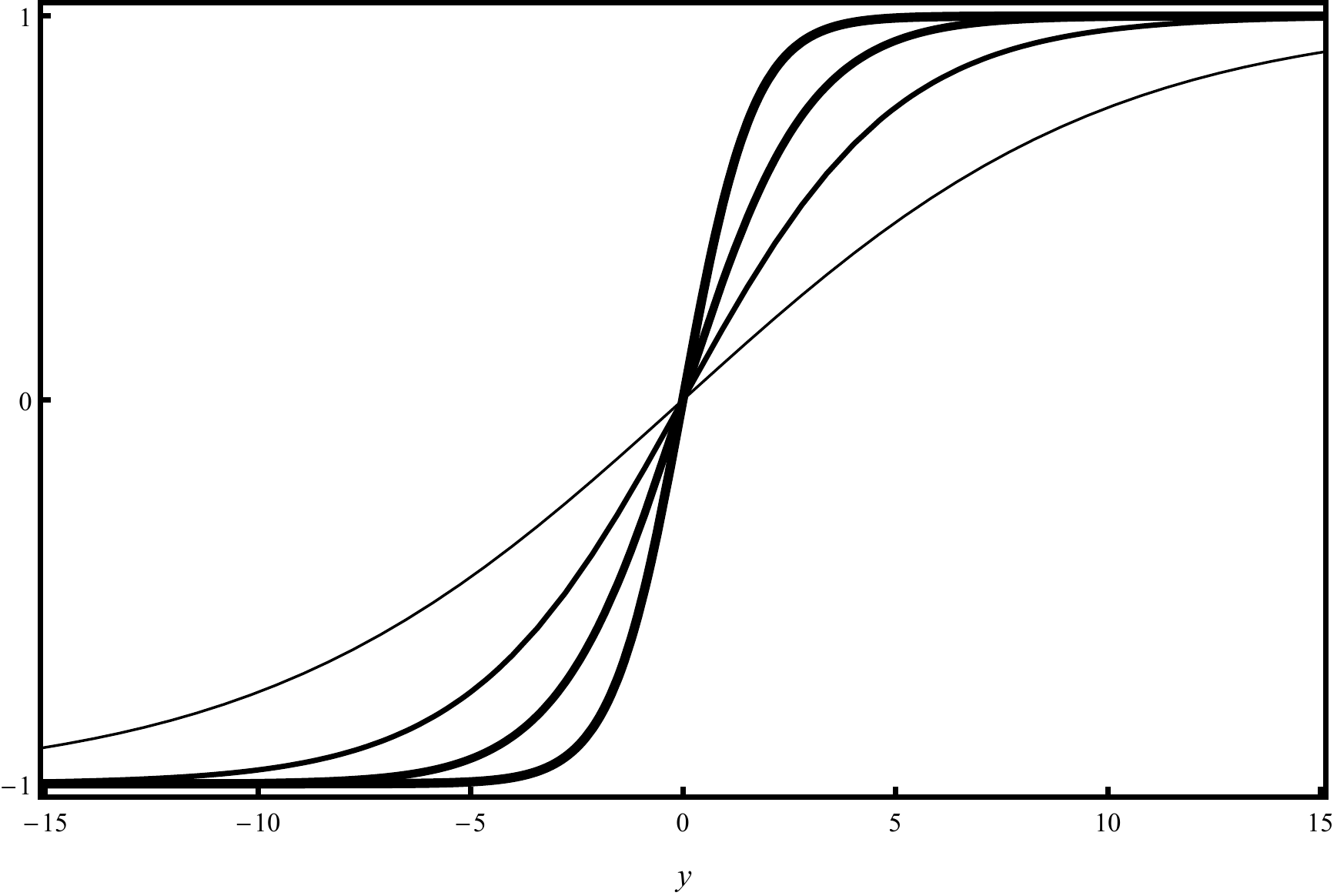} }
\subfigure[~Warp factor] 
{\label{fig1b}\includegraphics[width=7.2cm]{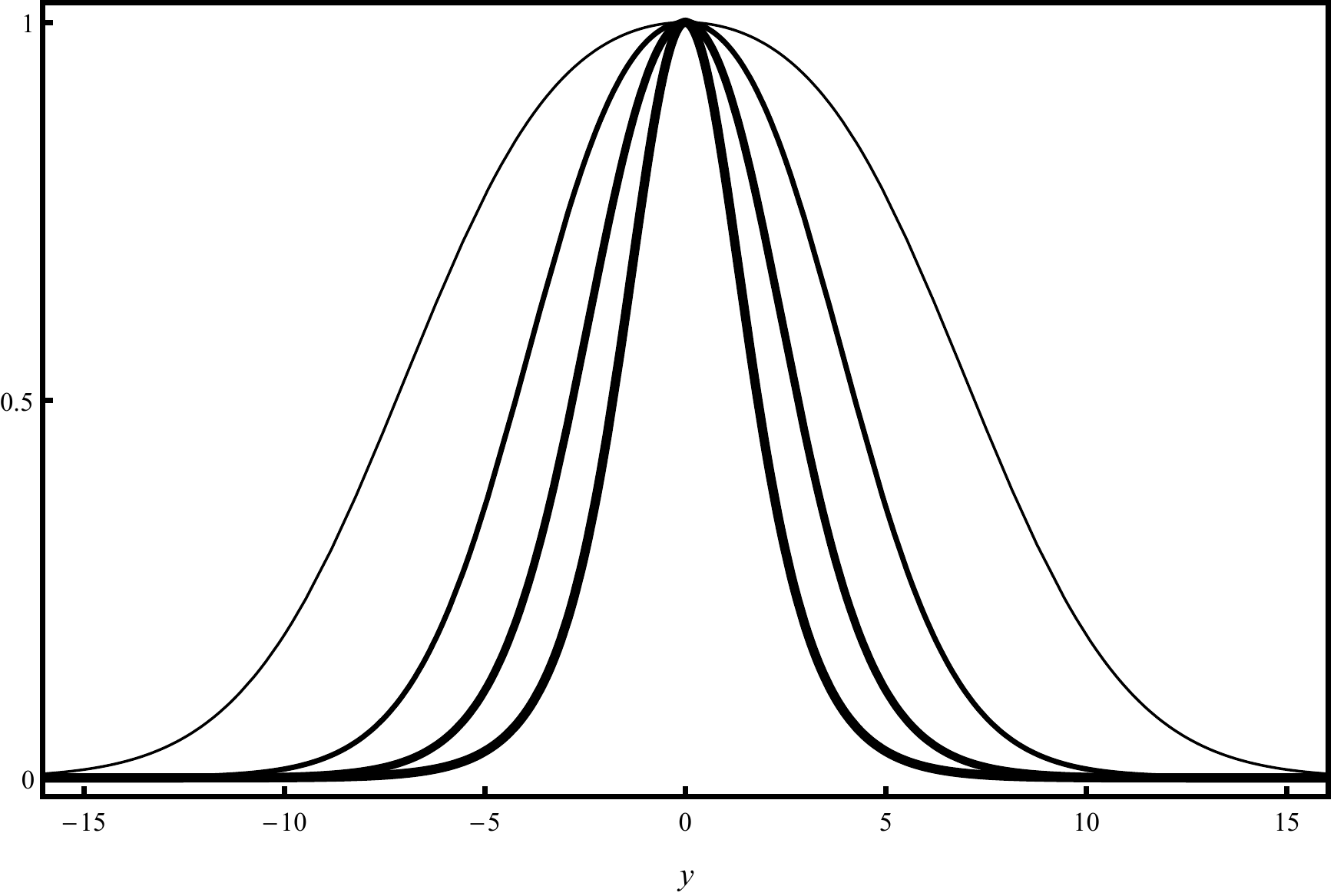} }
\subfigure[~Energy density] 
{\label{fig1c}\includegraphics[width=7.4cm]{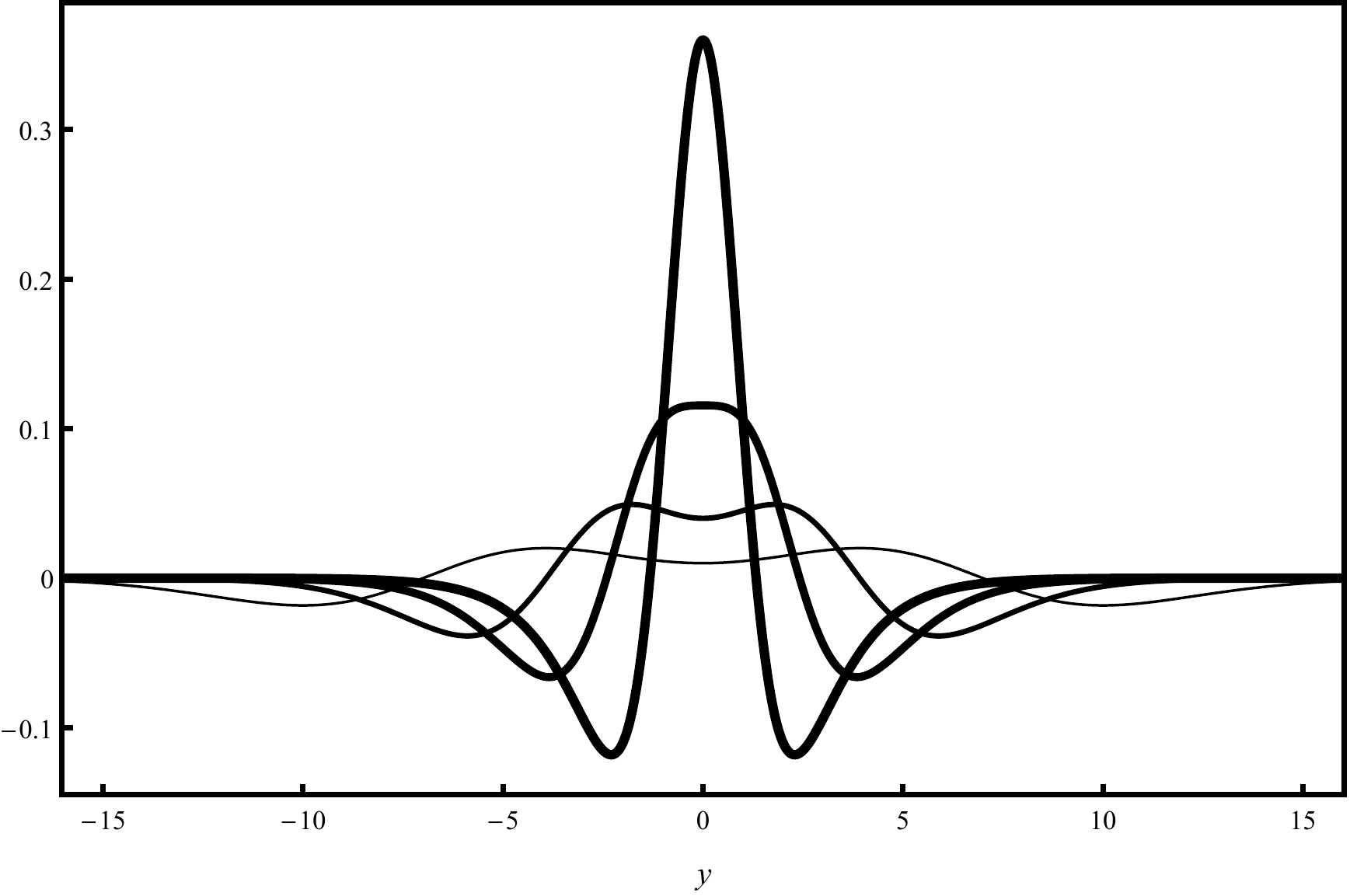}}
\caption{From top to bottom, we depict the solution $\phi(y)$, the warp factor and the energy density, respectively, associated to the model in Sec. \ref{sub1} with $c=0$. The parameter values are $r=0.05$, $r=0.1$, $r=0.17$ and $r=0.3$. The line thickness increases as $r$ increases. See the text for more details.}
\label{Fig1}
\end{figure}

The scalar field is governed by Eq. \eqref{phifo}, so the function given in Eq. \eqref{w1} provides the kinklike solution
\be\label{phi1}
\phi(y)=\tanh(2ry).
\ee
The profile of this solution is depicted in Fig.~\ref{fig1a}. Note that the parameter $r$ controls the thickness of the solution. In this case, we can combine the function in Eq. \eqref{W1} with the solution above to write Eq. \eqref{warpfo} in the form
\be\label{warp1}
A^\prime=-\frac23 \left[2r\,\tanh(2ry)-2\left(r-\frac{1}{3}\right)\tanh^{3}(2ry)+c\right],
\ee
whose integration leads to the following warp function
\be
A(y)=\frac{1}{9r}\left[(1-3r)\tanh(2ry)-2(\ln \cosh(2ry)+3cy)\right].
\ee

\noindent As usual, we used the condition $A(0)=0$. If we assume that $c=0$ and $r\in (0,1/2)$, the expression above describes precisely the warp function of the Bloch brane~\cite{Bazeia:2004dh}, which is generated from two coupled scalar fields. Thus, our model can mimic the Bloch brane using a single scalar field. For this case, we depict in Figs.~\ref{fig1b} and ~\ref{fig1c}, the warp factor and the energy density of the model, respectively. Note that in the interval $0.17\leq r<0.5$ the energy density has a single peak; however, for $r_{\rm crit}\approx 0.17$ a plateau appears. In the interval $0<r<r_{\rm crit}$ the energy density develops the two-hump behavior, engendering an internal structure.

Another important feature of this model appears when we consider $c\neq0$. As pointed out in Refs.~\cite{Bazeia:2013usa,Ahmed:2013mea,Bazeia:2019avw}, the addition of a constant in the $W$ function allows us to build asymmetric thick brane structures, where the five-dimensional cosmological constant assumes different values on each side of the brane. Based on this, we have checked that in the limit $y \rightarrow \pm \infty$ the cosmological constant provides the following result
\be \label{cosm1}
\Lambda_{5\pm}=-\frac{4}{3}\left[c+W(\phi(\pm\infty))\right]^{2}=-\frac{4}{3}\left(c\pm \frac{2}{3}\right)^{2}\,.
\ee 
Note that the cosmological constant is independent of the parameter $r$. In order to satisfy the condition in Eq. \eqref{normcond}, $c$ must be constrained to vary in the interval $(-2/3,2/3)$. In this interval the brane asymptotically connects two ${\rm AdS_{5}}$ geometries with different cosmological constants. 

\begin{figure}[htb!]
\centering
\subfigure[~Warp factor] {\label{fig2a}%
\includegraphics[width=7.2cm]{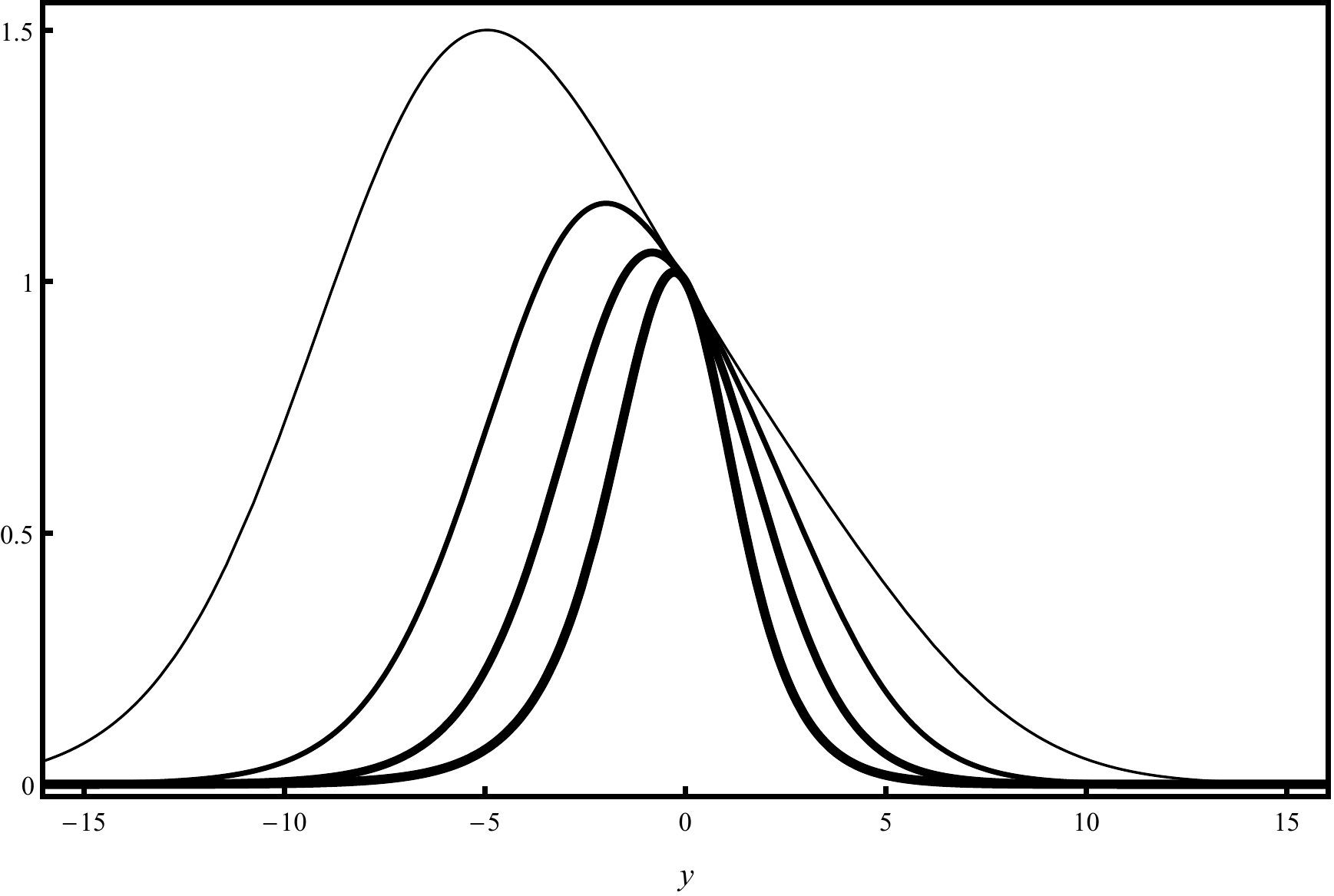} }
\subfigure[~Energy density] {\label{fig2b}
\includegraphics[width=7.4cm]{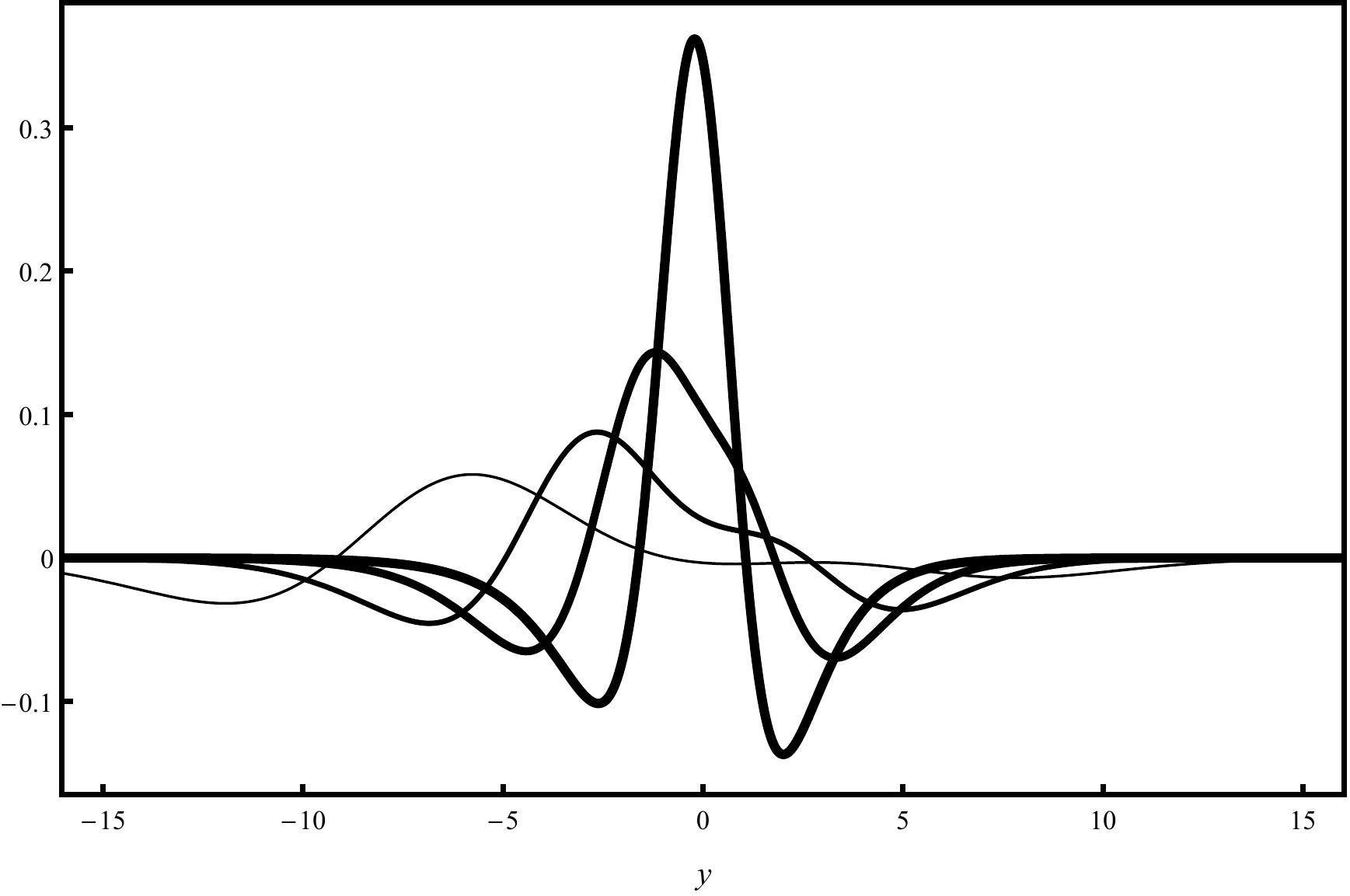} }
\caption{From top to bottom, we depict the warp factor and the energy density associated to the model in Sec. \ref{sub1}. We use the same conventions as Fig.~\ref{Fig1} with $c = 0.1$. See the text for more details.}
\label{Fig2}
\end{figure}

In Fig.~\ref{Fig2} we show the warp factor and the energy density of the model for a specific value of $c$, where we can see how the parameter modifies the profile of these quantities. Note that the warp factor does not have a maximum at $y=0$. Instead, it is shifted for a position $y_{\rm max}$, distancing from the origin as the parameter $r$ decreases. Here we recall that the asymmetry of the brane may be studied with the motivations to investigate the cosmic acceleration of the Universe \cite{Padilla:2004tp,Padilla:2004mc}. Also, in Ref. \cite{D}, the authors discussed a holographic description of an asymmetric version of the Randall-Sundrum braneworld model [2]. There, it was shown that that asymmetry induces a new characteristic in the massive KK spectrum, which is the appearance of a resonance state. In this sense, it is also natural to extend such study to the case of asymmetric thick brane considered in the present work. 
 
\subsection{Second model} \label{sub2}

The second model is described by the pair of functions
\be\label{w2}
\omega(\phi)=\alpha\left(\phi-\frac{\phi^{3}}{3}\right),
\ee
and
\be\label{W2}
W(\phi)=\tanh[\beta(\phi)]-\frac{1}{3}\tanh^{3}[\beta(\phi)]+\omega(\phi)+c\,,
\ee
with 
\be
\beta(\phi)=\frac{1}{\alpha}\left[\tanh^{-1}(\phi)-\phi\right].
\ee

\begin{figure}[htb!]
\centering
\subfigure[~Scalar field solution $\phi(y)$] 
{\label{fig3a}\includegraphics[width=7.2cm]{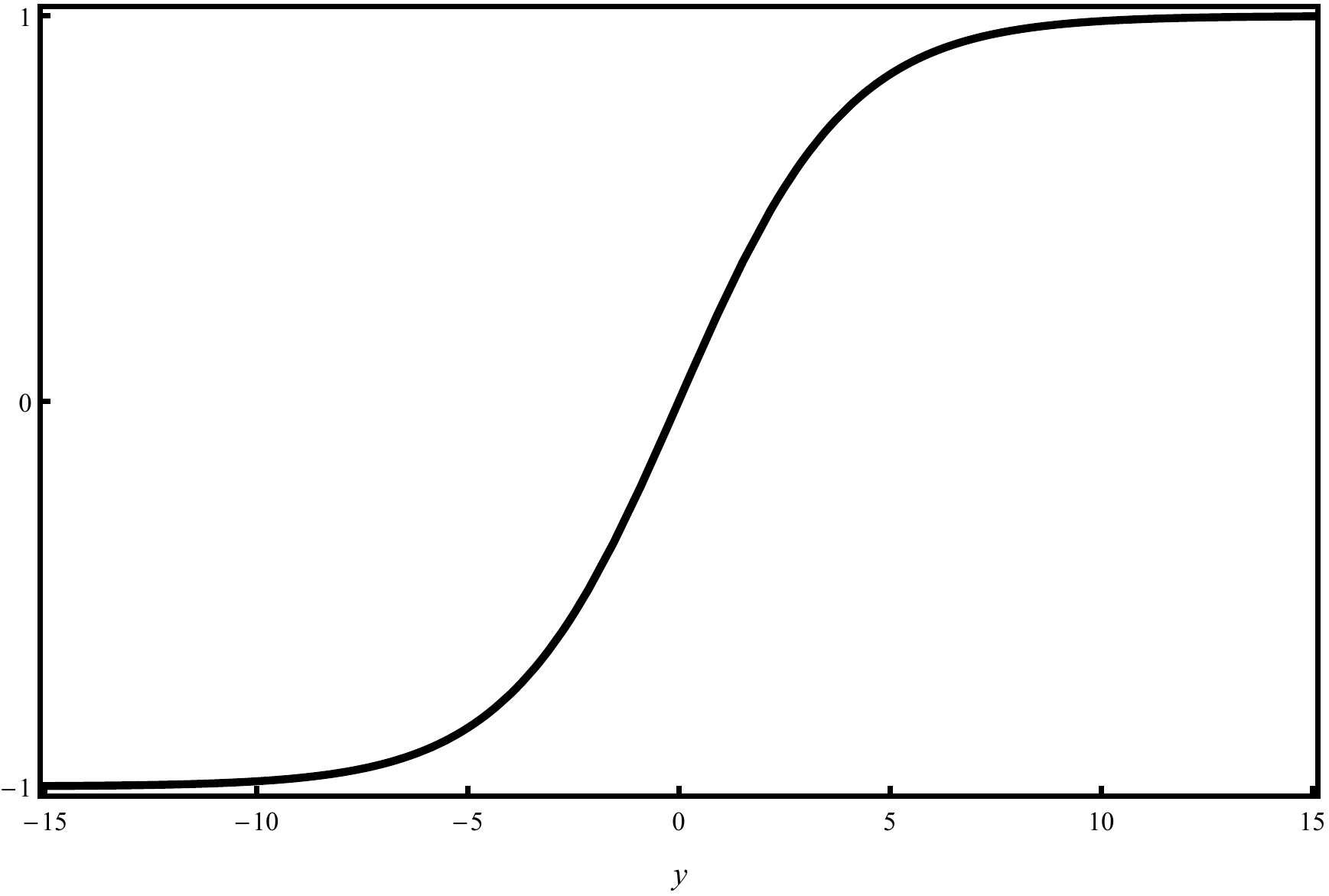} }
\subfigure[~Warp factor] 
{\label{fig3b}\includegraphics[width=7.2cm]{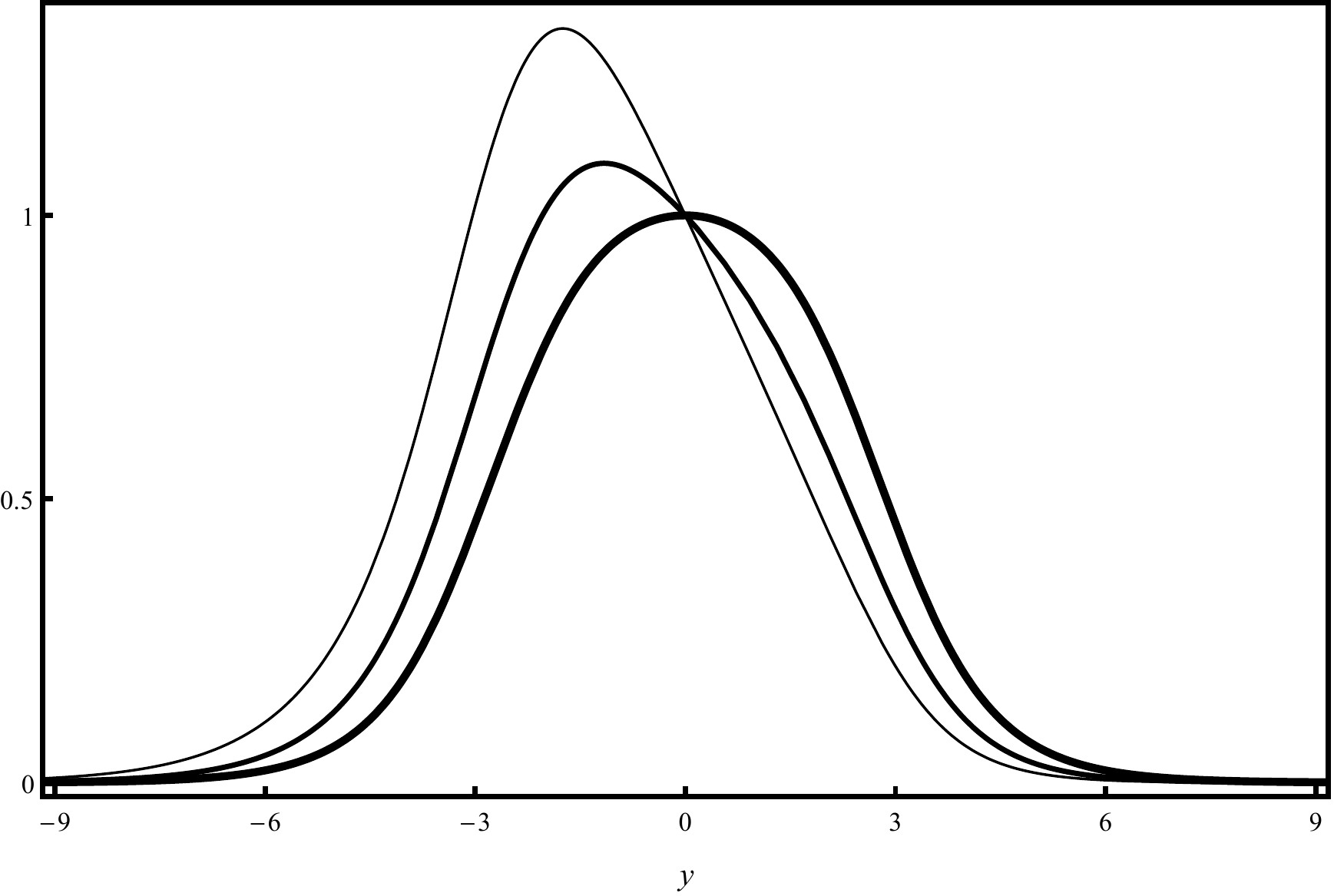} }
\subfigure[~Energy density] 
{\label{fig3c}\includegraphics[width=7.4cm]{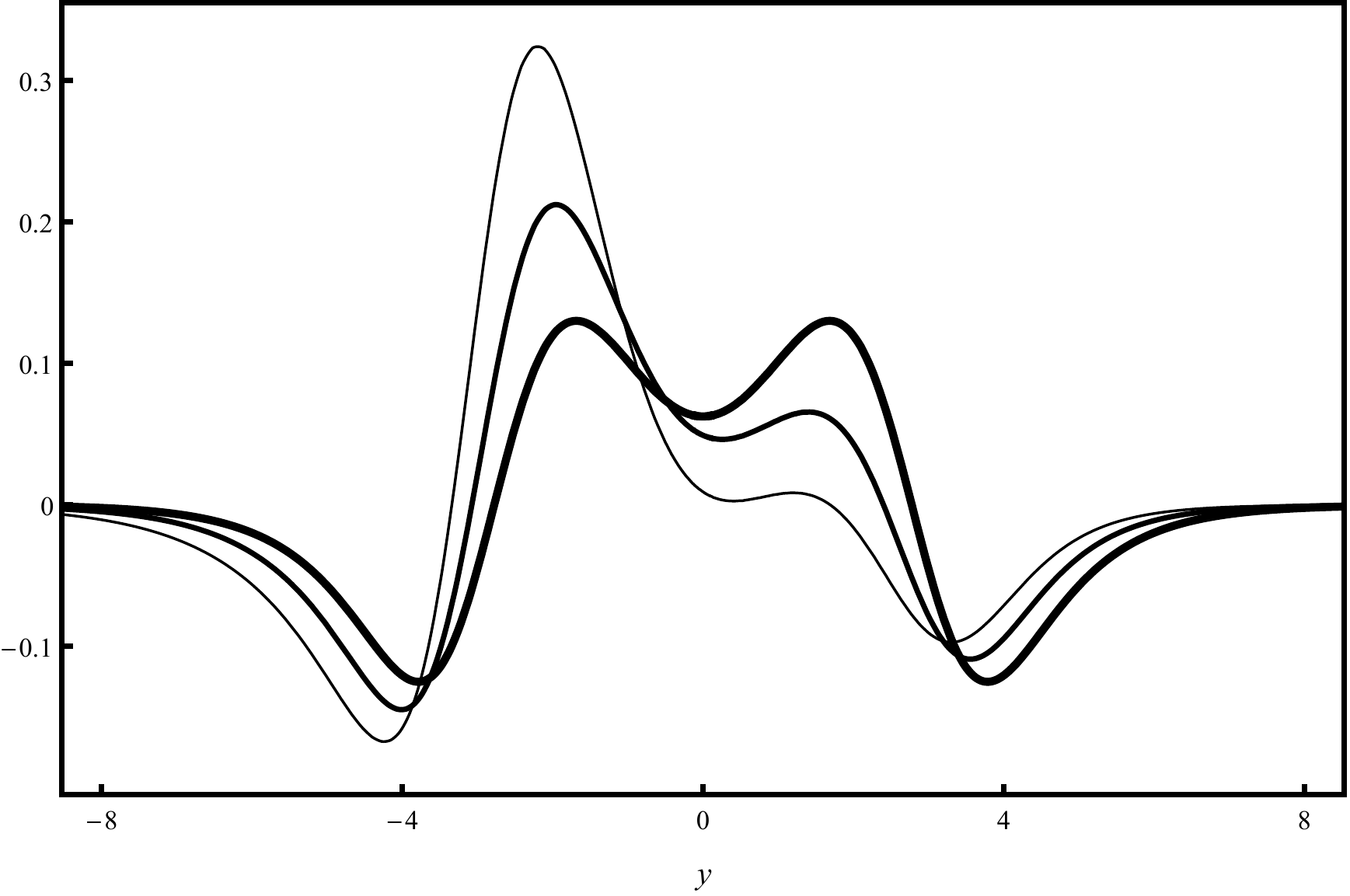}}
\caption{From top to bottom, we depict the solution $\phi(y)$, the warp factor and the energy density associated to the model in Sec. \ref{sub2}. We use $c=0$, $c=0.1$ and $c=0.2$ with $\alpha=1/4$. The line thickness decreases as $c$ increases. See the text for more details.}
\label{Fig3}
\end{figure}

\noindent Here, $\alpha$ and $c$ are real parameters. For this model, Eq. \eqref{phifo} allows us to obtain the kink-like solution
\be\label{phi1b}
\phi(y)=\tanh(\alpha y),
\ee
which is plotted in Fig.~\ref{fig3a}. We can use this solution to obtain the function $W(\phi)$ in terms of the extra dimension, i.e., $W(y)$ and then write the first order equation \eqref{warpfo} in the form
\be\label{warp2}
\begin{aligned}
	A^\prime &=-\frac23\bigg\{\tanh[\beta(y)]-\frac{\tanh^{3}[\beta(y)]}{3} \\
	         &\hspace{3.9mm}+\alpha\bigg[\tanh(\alpha y)-\frac{\tanh^{3}(\alpha y)}{3}\bigg]+c\bigg\}, 
\end{aligned}
\ee
where $\beta(y)=y-\tanh(\alpha y)/\alpha$. For $c=0$ the above equation is the same as that obtained in Ref.~\cite{Bazeia:2020qxr}, which is generated by two scalar fields with non-standard dynamics. This model is interesting because it engenders an internal structure that appears for $\alpha \leq \alpha_{\rm crit} \approx 1/2$. As we have done in the previous model, we add the constant $c$ to build asymmetric thick branes. 

Given the complexity of Eq. \eqref{warp2}, analytic solutions for the warp function $A(y)$ are not attainable. Before solving this equation numerically, we need to examine the interval for $c$, which gives rise to robust braneworld scenarios. In the limit $y \rightarrow \pm \infty$, we find that the five-dimensional cosmological constant takes the form
\be \label{cosm2}
\Lambda_{5\pm}=-\frac{4}{3}\left[c\pm \frac{2}{3}(\alpha+1)\right]^{2}\,.
\ee
This reveals that the interval $|c|\in (0,2/3(\alpha+1))$ allows us to build asymmetric thick branes, where the brane connects two ${\rm AdS_{5}}$ geometries with different cosmological constants. Since the interval depends on the parameter $\alpha$, we find the warp function numerically by using $\alpha < \alpha_{\rm crit}$ in addition to the condition $A(0)=0$. We made this choice for $\alpha$ to see how the asymmetry works to modify the internal structure of the brane. In Fig.~\ref{fig3b} and \ref{fig3c}, we depict the warp factor and energy density, respectively. Note that for $c=0$ the warp factor has a maximum in the position $ y=0 $, but for $c\neq 0$ it is shifted by a position $y_ {\rm max}$, distancing from the origin as $c$ increases. Note also that the parameter $c$ significantly modifies the energy density profile of the system. The increase of $c$ in this case is very important, that is, small variations of $c$ induces very large modifications in the asymmetric profile of the energy density, and this may be further studied concerning the asymmetry of the brane and its relation with the cosmic acceleration and with possible infrared modification of gravity, as explored before in Refs. \cite{Padilla:2004tp,Padilla:2004mc}, for instance. Moreover, it is also of interest concerning the hierarchy problem \cite{Dutra:2013jea}, new mechanisms for gravity localization \cite{Dvali:2000rv,Csaki:2000pp} and the presence of new resonance within the massive KK spectrum, as discussed before in Ref. \cite{D}. 

\subsection{Third model} \label{sub3}

\begin{figure}[htb!]
\centering
\subfigure[~Scalar field solution $\phi(y)$] {\label{fig4a}%
\includegraphics[width=7.2cm]{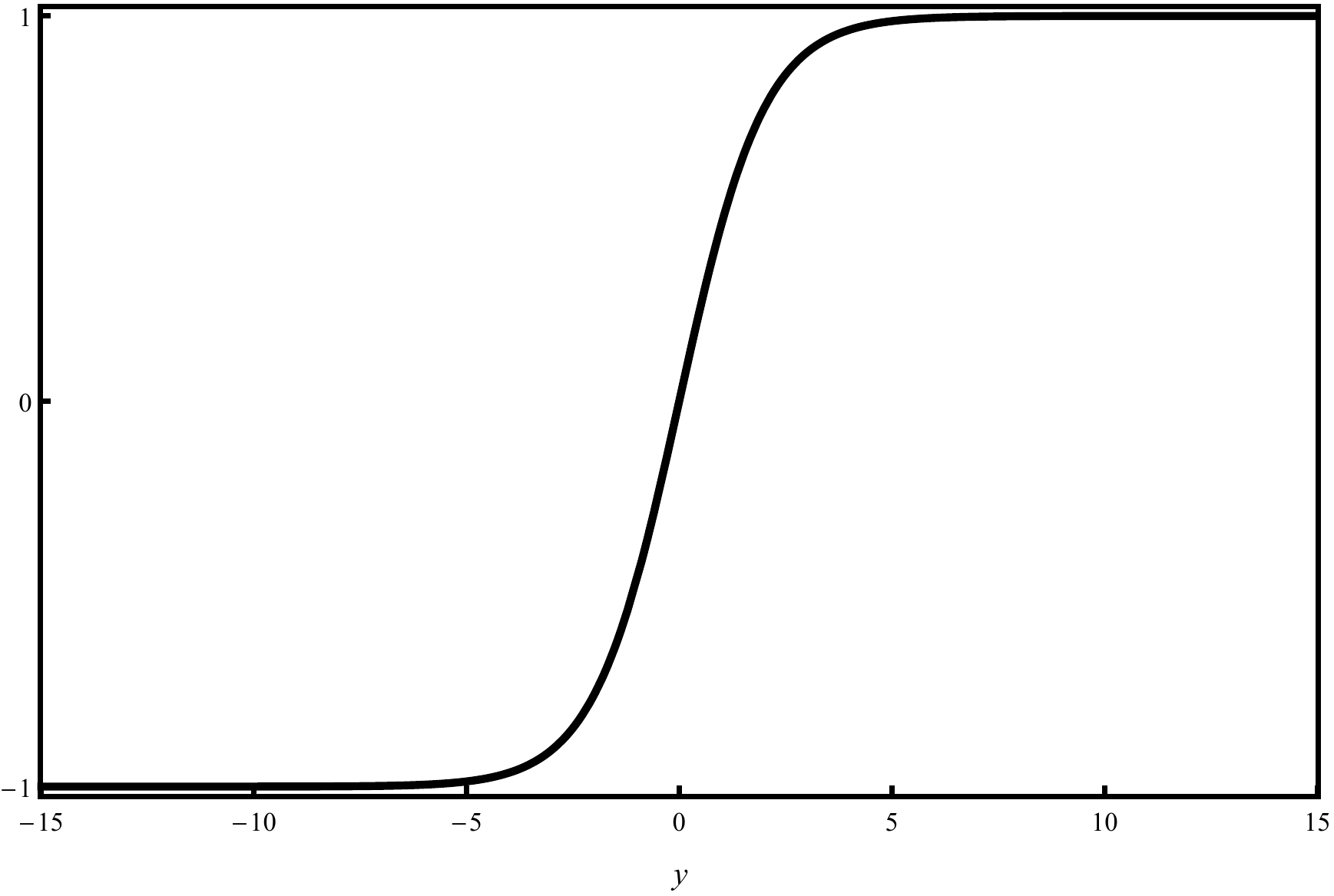} }
\subfigure[~Warp factor] 
{\label{fig4b}\includegraphics[width=7.2cm]{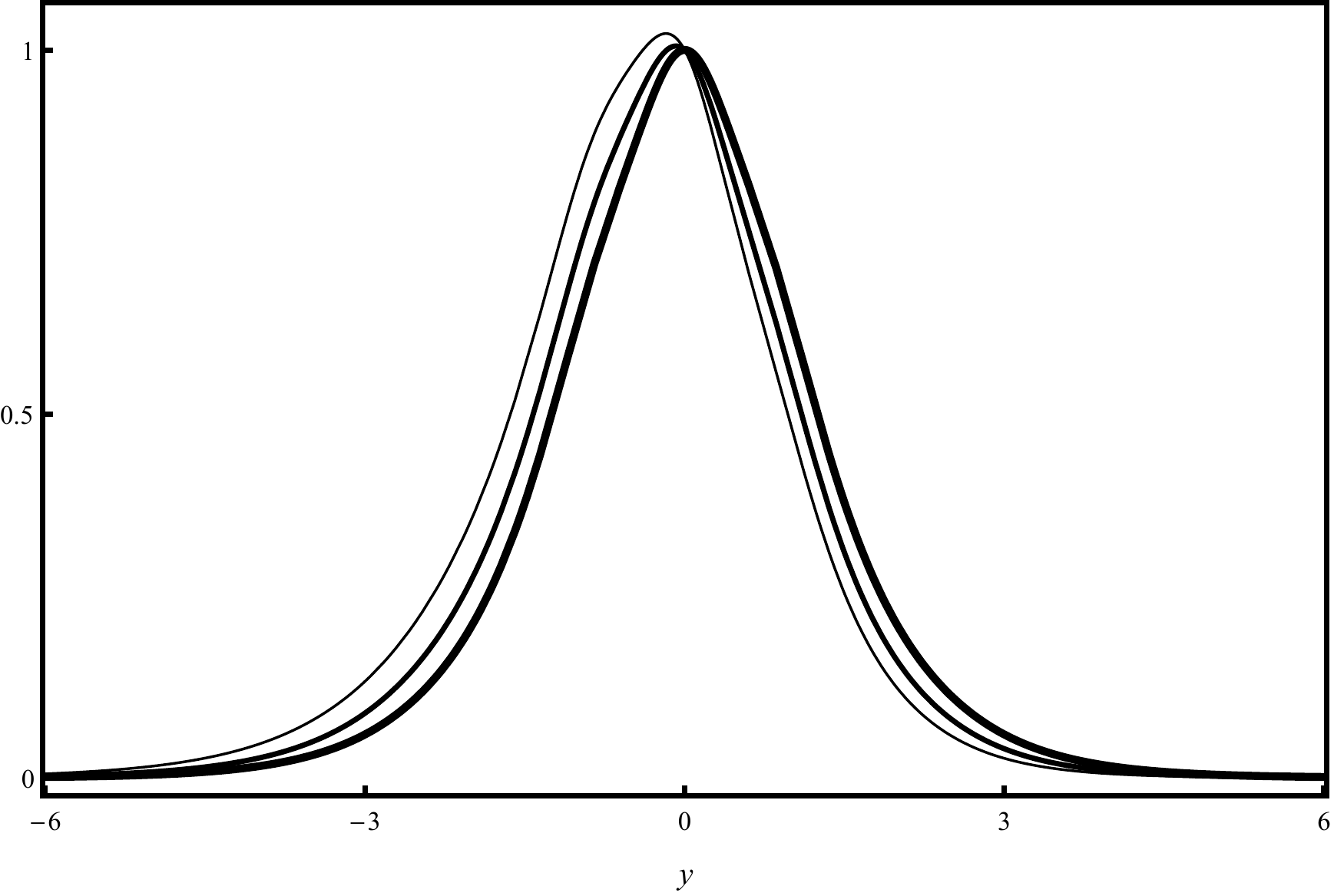} }
\subfigure[~Energy density] 
{\label{fig4c}\includegraphics[width=7.4cm]{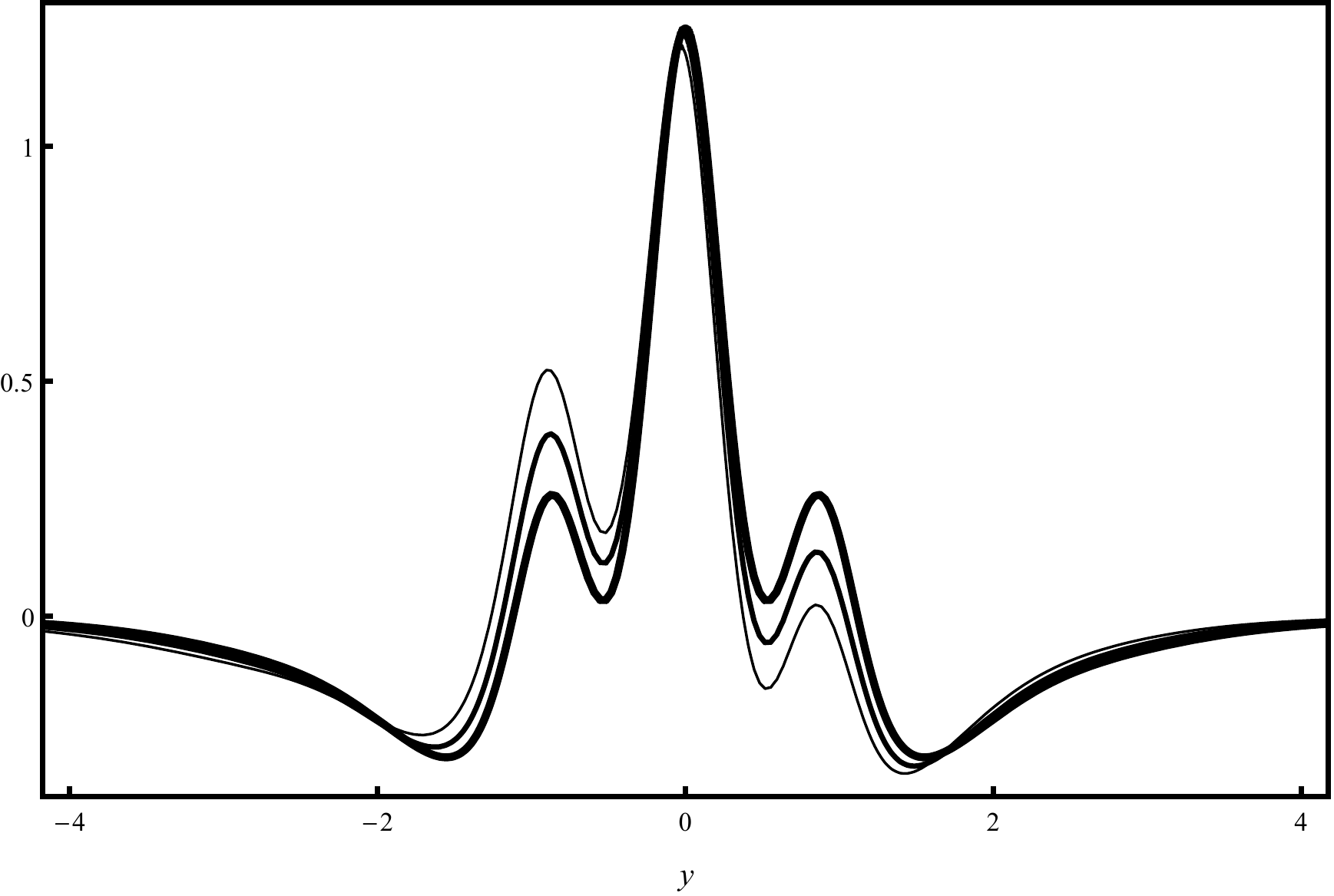}}
\caption{From top to bottom, we depict the solution $\phi(y)$, the warp factor and the energy density associated to the model in Sec. \ref{sub3}. We use $c=0$, $c=0.1$ and $c=0.2$ with $\alpha=1/2$ and $n=2$. The line thickness decreases as $c$ increases. See the text for more details.}
\label{Fig4}
\end{figure}

The third model is described by the same $\omega(\phi)$ of the previous model, in Eq. \eqref{w2}, but here we choose another $W(\phi)$, of the form
\be
W(\phi)=\tanh[\gamma(\phi)]-\frac{1}{3}\tanh^{3}[\gamma(\phi)]+\omega(\phi)+c\,,
\ee
with
\be
\gamma(\phi)=\frac{\tanh^{-1}(\phi)}{2\alpha}+\frac{1}{4\alpha}\left[\Ci(\xi^{+}_{n}(\phi))-\Ci(\xi^{-}_{n}(\phi))\right],
\ee
and $\xi^{\pm}_{n}(\phi)=2n\pi(1\pm \phi)$, with $ n \in \mathbb{N} $. As we have already discussed, the function $\omega(\phi)$ allows us to obtain the solution $\phi(y)=\tanh(\alpha y)$. 
The cosine integral function, $\Ci(z)$, with the argument $z$ is defined as $\Ci(z)=-\int_z^{\infty} \left( \cos x/x \right)\, dx = \gamma+\ln(z)+\int_0^z \left[ (\cos (y) -1)/y  \right] \,dy$, and $\gamma=0.577$ is the Euler-Mascheroni constant.
By combining this solution with the function $W(\phi)$ above, we can write Eq. \eqref{warpfo} as
\be\label{warp3}
\begin{aligned}
	A^\prime &=-\frac23\Bigg\{\tanh[\gamma(y)]-\frac{\tanh^{3}[\gamma(y)]}{3} \\
	         &\hspace{3.9mm}+\alpha\bigg[\tanh(\alpha y)-\frac{\tanh^{3}(\alpha y)}{3}\bigg]+c\Bigg\}, 
\end{aligned}
\ee
with
\be
\gamma(y)=\frac{y}{2}+\frac{1}{4\alpha}\left[\Ci(\xi^{+}_{n}(y))-\Ci(\xi^{-}_{n}(y))\right],
\ee
and $\xi^{\pm}_{n}(y)=2n\pi[1\pm \tanh(\alpha y)]$. This result was also obtained in Ref.~\cite{Bazeia:2020qxr}, in the presence of two scalar fields with non-standard dynamics in flat spacetime. However, it was shown that Eq.~\eqref{warp3} appears due to a multikink structure that appears in one of the fields \cite{Bazeia:2019vld}, while we reproduce the same model using a single scalar field with a usual kink-like structure.

For this model we have examined the five-dimensional cosmological constant and found that it can be expressed in the same way as Eq. \eqref{cosm2}. We then choose $\alpha=1/2$ and $n=2$, to solve Eq. \eqref{warp3} numerically for some values of $c$. In Fig.~\ref{Fig4} we plot the solution, the warp factor and the energy density of the model. Note that for $c=0$ the warp factor has a maximum in the position $ y=0 $, but for $c\neq 0$ it is slightly shifted for a position $y_ {\rm max}$ distancing from the origin as $c$ increases. As commented before, the asymmetry identified in this model may be of interest concerning the acceleration of the Universe [26,27], the hierarchy problem \cite{Dutra:2013jea}, new mechanisms for gravity localization \cite{Dvali:2000rv,Csaki:2000pp} and the induction of resonance inside the massive KK spectrum, as discussed before in Ref. \cite{D}.

We have also checked that as the parameter $\alpha$ increases, the warp factor becomes thinner and the energy density becomes more concentrated around the origin. In addition, the parameter $n$ plays an important role in the formation of internal structures, controlling the number of peaks in the energy density of the system. 

\section{Conclusion} \label{sec:Conclusion}

In this work we studied the construction of thick brane structures in the warped five-dimensional braneworld scenario with a single extra spatial dimension of infinite extent. We considered a specific class of modified theories of gravity in the presence of a Lagrange multiplier, as recently studied in \cite{Zhong:2017uhn,Bazeia:2020zut}. In the last work \cite{Bazeia:2020zut}, an interesting first order framework was developed, which helped us to describe analytic solutions that are robust against fluctuations in the metric. We used this first order framework in the present work, to construct new braneworld solutions, which engender profiles that are similar to braneworld solutions obtained before in the case of standard gravity in the presence of two scalar fields, but here we considered models described by a single scalar field. We illustrated this possibility with several distinct models, which show that the procedure described in this work is robust and can be used to describe new results of current interest. 

In the case of thick braneworld scenarios in the presence of standard gravity, featuring two scalar fields, the second field may be used to modify the internal profile of the brane. Here, however, we have shown that the presence of the Lagrange multiplier allows the inclusion of another auxiliary function, $W=W(\phi)$, which, even in the absence of the second field, is capable of modifying the internal structure of the brane. 
This result is of current interest, since it allows that we modify the internal structure of the brane even in the absence of the second scalar field. In this sense, the present work unveils another route to induce internal structure to the brane, via the presence of the Lagrange multiplier. This is new and somehow easy to implement, within the context of the first order framework developed in \cite{Bazeia:2020qxr} and further considered in the present work. 

The main results of this work can be used in several distinct directions, in particular in the study of gravitational resonances which was recently considered in \cite{Zhong:2018fdq}. They may also suggest the study concerning the entrapment of scalar, spinor and gauge fields inside the brane. This subject has been studied before in a diversity of situations, but no investigation in the specific case in the presence of Lagrange multipliers has been done yet. In particular, we notice from the results discussed in \cite{Almeida:2009jc} that branes with internal structures seem to be more effective to entrap matter, in comparison with branes without internal structures. This motivates us to study the entrapment of matter in this new scenario, in the presence of Lagrange multipliers. 

In the same line, the localization of gauge and Kalb-Ramond fields in branes usually requires the addition of an extra scalar field, the dilaton \cite{Kehagias:2000au}. Thus, in the braneworld scenario with standard gravity and two scalar fields, the appearance of another scalar, the dilaton, leads us with three scalar fields, and this makes the investigation harder to understand; see, e.g., Ref. \cite{Cruz:2012kd}. In this sense, investigations concerning localization of gauge and Kalb-Ramond fields in braneworld models with Lagrange multipliers may shed new light into the subject. Another line of research concerns the addition of scalar field with the cuscuton kinematics. This has been studied before in \cite{Bazeia:2012br,Andrade:2018afh} and we think that the cuscuton modification may contribute to deform the warp factor and then the gravitational sector of the brane. Another issue concerns investigating corrections of the Newton’s law of gravitation due to modifications added in the present work, to see how the Newtonian limit of these new models conform with the phenomenology of branes. This possibility may be implemented under the lines of Ref. \cite{Veras:2017nke}. These and other related issues are presently under consideration, and we hope to report them in the near future.

\section*{Acknowledgments}
DB is supported by Conselho Nacional de Desenvolvimento Cient\'\i fico e Tecnol\'ogico, grants No. 404913/2018-0 and No. 303469/2019-6, and by Paraiba State Research Foundation, grant No. 0015/2019.
DB and DAF would like to thank CAPES, CNPq and Para\'\i ba State Research Foundation (FAPESQ/PB, Grant 0015/2019) for partial financial support.
FSNL acknowledges support from the Funda\c{c}\~{a}o para a Ci\^{e}ncia e a Tecnologia (FCT) Scientific Employment Stimulus contract with reference CEECIND/04057/2017, and the FCT research grants No. UID/FIS/04434/2020, No. PTDC/FIS-OUT/29048/2017 and No. CERN/FIS-PAR/0037/2019.
JLR is supported by the European Regional Development Fund and the programme Mobilitas Pluss (MOBJD647).



\begin{thebibliography}{99}

\bibitem{Randall:1999ee}
L.~Randall and R.~Sundrum,
``A Large mass hierarchy from a small extra dimension,''
Phys. Rev. Lett. \textbf{83} (1999), 3370-3373
[arXiv:hep-ph/9905221 [hep-ph]].

\bibitem{Randall:1999vf}
L.~Randall and R.~Sundrum,
``An Alternative to compactification,''
Phys. Rev. Lett. \textbf{83} (1999), 4690-4693
[arXiv:hep-th/9906064 [hep-th]].

\bibitem{Goldberger:1999uk}
W.~D.~Goldberger and M.~B.~Wise,
``Modulus stabilization with bulk fields,''
Phys. Rev. Lett. \textbf{83} (1999), 4922-4925
[arXiv:hep-ph/9907447 [hep-ph]].

\bibitem{DeWolfe:1999cp}
O.~DeWolfe, D.~Z.~Freedman, S.~S.~Gubser and A.~Karch,
``Modeling the fifth-dimension with scalars and gravity,''
Phys. Rev. D \textbf{62} (2000), 046008
[arXiv:hep-th/9909134 [hep-th]].

\bibitem{Gremm:1999pj}
M.~Gremm,
``Four-dimensional gravity on a thick domain wall,''
Phys. Lett. B \textbf{478} (2000), 434-438
[arXiv:hep-th/9912060 [hep-th]].

\bibitem{Csaki:2000fc}
C.~Csaki, J.~Erlich, T.~J.~Hollowood and Y.~Shirman,
``Universal aspects of gravity localized on thick branes,''
Nucl. Phys. B \textbf{581} (2000), 309-338
[arXiv:hep-th/0001033 [hep-th]].

\bibitem{Brito:2001hd}
F.~Brito, M.~Cvetic and S.~Yoon,
``From a thick to a thin supergravity domain wall,''
Phys. Rev. D \textbf{64} (2001), 064021
[arXiv:hep-ph/0105010 [hep-ph]].

\bibitem{Kehagias:2000au}
A.~Kehagias and K.~Tamvakis,
``Localized gravitons, gauge bosons and chiral fermions in smooth spaces generated by a bounce,''
Phys. Lett. B \textbf{504} (2001), 38-46
[arXiv:hep-th/0010112 [hep-th]].

\bibitem{Campos:2001pr}
A.~Campos,
``Critical phenomena of thick branes in warped space-times,''
Phys. Rev. Lett. \textbf{88} (2002), 141602
[arXiv:hep-th/0111207 [hep-th]].

\bibitem{Bazeia:2003aw}
D.~Bazeia, C.~Furtado and A.~R.~Gomes,
``Brane structure from scalar field in warped space-time,''
JCAP \textbf{02} (2004), 002
[arXiv:hep-th/0308034 [hep-th]].

\bibitem{Bazeia:2004dh}
D.~Bazeia and A.~R.~Gomes,
``Bloch brane,''
JHEP \textbf{05} (2004), 012
[arXiv:hep-th/0403141 [hep-th]].

\bibitem{deSouzaDutra:2008gm}
A.~de Souza Dutra, A.~C.~A.~de Faria, Jr. and M.~Hott,
``Degenerate and critical Bloch branes,''
Phys. Rev. D \textbf{78} (2008), 043526
[arXiv:0807.0586 [hep-th]].

\bibitem{Dutra:2013jea}
A.~de Souza Dutra, G.~P.~de Brito and J.~M.~Hoff da Silva,
``Asymmetrical bloch branes and the hierarchy problem,''
EPL \textbf{108} (2014) no.1, 11001
[arXiv:1312.0091 [hep-th]].

\bibitem{Bazeia:2016uhr}
D.~Bazeia, E.~E.~M.~Lima and L.~Losano,
``Hybrid Bloch Brane,''
Eur. Phys. J. C \textbf{77} (2017) no.2, 127
[arXiv:1611.09314 [hep-th]].


\bibitem{Xie:2015dva}
Q.~Y.~Xie, H.~Guo, Z.~H.~Zhao, Y.~Z.~Du and Y.~P.~Zhang,
``Spectrum structure of a fermion on Bloch branes with two scalar\textendash{}fermion couplings,''
Class. Quant. Grav. \textbf{34} (2017) no.5, 055007
[arXiv:1510.03345 [hep-th]].
\bibitem{Almeida:2018bzx}
C.~A.~S.~Almeida, D.~F.~S.~Veras and D.~M.~Dantas,
``Corrections to Newton\textquoteright{}s law of gravitation - application to hybrid Bloch brane,''
J. Phys. Conf. Ser. \textbf{965} (2018) no.1, 012002
[arXiv:1802.06808 [gr-qc]].

\bibitem{Brito:2019bbb}
F.~A.~Brito, L.~Losano and J.~R.~L.~Santos,
``The Extension Method for Bloch Branes,''
[arXiv:1911.00191 [hep-th]].

\bibitem{Almeida:2009jc}
C.~A.~S.~Almeida, M.~M.~Ferreira, Jr., A.~R.~Gomes and R.~Casana,
``Fermion localization and resonances on two-field thick branes,''
Phys. Rev. D \textbf{79} (2009), 125022
[arXiv:0901.3543 [hep-th]].

\bibitem{Castro:2010au}
L.~B.~Castro,
``Fermion localization on two-field thick branes,''
Phys. Rev. D \textbf{83} (2011), 045002
[arXiv:1008.3665 [hep-th]].

\bibitem{Cruz:2013zka}
W.~T.~Cruz, R.~V.~Maluf and C.~A.~S.~Almeida,
``Kalb-Ramond field localization on the Bloch brane,''
Eur. Phys. J. C \textbf{73} (2013), 2523
[arXiv:1303.1096 [hep-th]].


\bibitem{Cruz:2012kd}
W.~T.~Cruz, A.~R.~P.~Lima and C.~A.~S.~Almeida,
``Gauge field localization on the Bloch Brane,''
Phys. Rev. D \textbf{87} (2013) no.4, 045018
[arXiv:1211.7355 [hep-th]].

\bibitem{Xie:2013rka}
Q.~Y.~Xie, J.~Yang and L.~Zhao,
``Resonance Mass Spectra of Gravity and Fermion on Bloch Branes,''
Phys. Rev. D \textbf{88} (2013), 105014
[arXiv:1310.4585 [hep-th]].

\bibitem{Zhao:2014gka}
Z.~H.~Zhao, Y.~X.~Liu and Y.~Zhong,
``U(1) gauge field localization on a Bloch brane with Chumbes-Holf da Silva-Hott mechanism,''
Phys. Rev. D \textbf{90} (2014) no.4, 045031
[arXiv:1402.6480 [hep-th]].

\bibitem{Bazeia:2020qxr}
D.~Bazeia, D.~A.~Ferreira and M.~A.~Marques,
``Symmetric and asymmetric thick brane structures,''
Eur. Phys. J. Plus \textbf{135} (2020) no.7, 587
[arXiv:2004.11398 [hep-th]].

\bibitem{Bazeia:2019vld}
D.~Bazeia, M.~A.~Liao and M.~A.~Marques,
``Geometrically constrained kinklike configurations,''
Eur. Phys. J. Plus \textbf{135}, no.4, 383 (2020)
[arXiv:1908.01085 [hep-th]].
\bibitem{Padilla:2004tp}
A.~Padilla,
``Cosmic acceleration from asymmetric branes,''
Class. Quant. Grav. \textbf{22} (2005), 681-694
[arXiv:hep-th/0406157 [hep-th]].

\bibitem{Padilla:2004mc}
A.~Padilla,
``Infra-red modification of gravity from asymmetric branes,''
Class. Quant. Grav. \textbf{22} (2005) no.6, 1087-1104
[arXiv:hep-th/0410033 [hep-th]].
 
\bibitem{Dutra:2014xla}
A.~de Souza Dutra, G.~P.~de Brito and J.~M.~Hoff da Silva,
``Method for obtaining thick brane models,''
Phys. Rev. D \textbf{91} (2015) no.8, 086016
[arXiv:1412.5543 [hep-th]].

\bibitem{Gregory:2000jc}
R.~Gregory, V.~A.~Rubakov and S.~M.~Sibiryakov,
``Opening up extra dimensions at ultra large scales,''
Phys. Rev. Lett. \textbf{84} (2000), 5928-5931
[arXiv:hep-th/0002072 [hep-th]].

\bibitem{Dvali:2000rv}
G.~R.~Dvali, G.~Gabadadze and M.~Porrati,
``Metastable gravitons and infinite volume extra dimensions,''
Phys. Lett. B \textbf{484} (2000), 112-118
[arXiv:hep-th/0002190 [hep-th]].

\bibitem{Csaki:2000pp}
C.~Csaki, J.~Erlich and T.~J.~Hollowood,
``Quasilocalization of gravity by resonant modes,''
Phys. Rev. Lett. \textbf{84} (2000), 5932-5935
[arXiv:hep-th/0002161 [hep-th]].
 
\bibitem{Bazeia:2013usa}
D.~Bazeia, R.~Menezes and R.~da Rocha,
``A Note on Asymmetric Thick Branes,''
Adv. High Energy Phys. \textbf{2014} (2014), 276729
[arXiv:1312.3864 [hep-th]].

\bibitem{Ahmed:2013mea}
A.~Ahmed, L.~Dulny and B.~Grzadkowski,
``Generalized Randall-Sundrum model with a single thick brane,''
Eur. Phys. J. C \textbf{74} (2014), 2862
[arXiv:1312.3577 [hep-th]].

\bibitem{Bazeia:2019avw}
D.~Bazeia and D.~A.~Ferreira,
``New results on asymmetric thick branes,''
Annals Phys. \textbf{411} (2019), 167975
[arXiv:1910.09530 [hep-th]].

\bibitem{Afonso:2007gc}
V.~I.~Afonso, D.~Bazeia, R.~Menezes and A.~Y.~Petrov,
``f(R)-Brane,''
Phys. Lett. B \textbf{658} (2007), 71-76
[arXiv:0710.3790 [hep-th]].

\bibitem{Dzhunushaliev:2009dt}
V.~Dzhunushaliev, V.~Folomeev, B.~Kleihaus and J.~Kunz,
``Some thick brane solutions in f(R)-gravity,''
JHEP \textbf{04} (2010), 130
[arXiv:0912.2812 [gr-qc]].

\bibitem{Zhong:2010ae}
Y.~Zhong, Y.~X.~Liu and K.~Yang,
``Tensor perturbations of $f(R)$-branes,''
Phys. Lett. B \textbf{699} (2011), 398-402
[arXiv:1010.3478 [hep-th]].

\bibitem{Liu:2012rc}
Y.~X.~Liu, K.~Yang, H.~Guo and Y.~Zhong,
``Domain Wall Brane in Eddington Inspired Born-Infeld Gravity,''
Phys. Rev. D \textbf{85} (2012), 124053
[arXiv:1203.2349 [hep-th]].

\bibitem{Bazeia:2013uva}
D.~Bazeia, A.~S.~Lob\~ao, Jr., R.~Menezes, A.~Y.~Petrov and A.~J.~da Silva,
``Braneworld solutions for F(R) models with non-constant curvature,''
Phys. Lett. B \textbf{729} (2014), 127-135
[arXiv:1311.6294 [hep-th]].
\bibitem{Menezes:2014bta}
R.~Menezes,
``First Order Formalism for Thick Branes in Modified Teleparallel Gravity,''
Phys. Rev. D \textbf{89} (2014) no.12, 125007
[arXiv:1403.5587 [hep-th]].

\bibitem{Bazeia:2014poa}
D.~Bazeia, L.~Losano, R.~Menezes, G.~J.~Olmo and D.~Rubiera-Garcia,
``Thick brane in $f(R)$ gravity with Palatini dynamics,''
Eur. Phys. J. C \textbf{75} (2015) no.12, 569
[arXiv:1411.0897 [hep-th]].

\bibitem{Bazeia:2015owa}
D.~Bazeia, A.~S.~Lob\~ao and R.~Menezes,
``Thick brane models in generalized theories of gravity,''
Phys. Lett. B \textbf{743} (2015), 98-103
[arXiv:1502.04757 [hep-th]].

\bibitem{Fu:2016rgr}
Q.~M.~Fu, L.~Zhao, Y.~Z.~Du and B.~M.~Gu,
``Resonances of Spin-1/2 Fermions in Eddington-Inspired Born-Infeld Gravity,''
Commun. Theor. Phys. \textbf{65} (2016) no.3, 292-300

\bibitem{Fu:2016szo}
Q.~M.~Fu, L.~Zhao, B.~M.~Gu, K.~Yang and Y.~X.~Liu,
``Hybrid metric-Palatini brane system,''
Phys. Rev. D \textbf{94} (2016) no.2, 024020
[arXiv:1601.06546 [gr-qc]].

\bibitem{Rosa:2020uli}
J.~L.~Rosa, D.~A.~Ferreira, D.~Bazeia and F.~S.~N.~Lobo,
Eur. Phys. J. C \textbf{81} (2021) no.1, 20
doi:10.1140/epjc/s10052-021-08840-3
[arXiv:2010.10074 [gr-qc]].

\bibitem{Mazani:2020abe}
E.~Mazani, A.~Tofighi and M.~M.~Sorkhi,
``Fermion and graviton in Dirac\textendash{}Born\textendash{}Infeld braneworld models,''
Eur. Phys. J. C \textbf{80} (2020) no.3, 267

\bibitem{Zhong:2017uhn}
Y.~Zhong, Y.~Zhong, Y.~P.~Zhang and Y.~X.~Liu,
``Thick branes with inner structure in mimetic gravity,''
Eur. Phys. J. C \textbf{78} (2018) no.1, 45
[arXiv:1711.09413 [hep-th]].

\bibitem{Bazeia:2020zut}
D.~Bazeia, D.~A.~Ferreira and D.~C.~Moreira,
``First order formalism for thick branes in modified gravity with Lagrange multiplier,''
EPL \textbf{129} (2020) no.1, 11004
[arXiv:2002.00229 [hep-th]].

\bibitem{A}
E.~A.~Lim, I.~Sawicki and A.~Vikman,
``Dust of Dark Energy,''
JCAP \textbf{05} (2010), 012
[arXiv:1003.5751 [astro-ph.CO]].


\bibitem{B}
S.~Capozziello, J.~Matsumoto, S.~Nojiri and S.~D.~Odintsov,
``Dark energy from modified gravity with Lagrange multipliers,''
Phys. Lett. B \textbf{693} (2010), 198-208
[arXiv:1004.3691 [hep-th]].


\bibitem{C}
C.~Gao, Y.~Gong, X.~Wang and X.~Chen,
``Cosmological models with Lagrange Multiplier Field,''
Phys. Lett. B \textbf{702} (2011), 107-113
[arXiv:1003.6056 [astro-ph.CO]].


\bibitem{D}
G.~Gabadadze, L.~Grisa and Y.~Shang,
``Resonance in asymmetric warped geometry,''
JHEP \textbf{08} (2006), 033
[arXiv:hep-th/0604218 [hep-th]].


\bibitem{Zhong:2018fdq}
Y.~Zhong, Y.~P.~Zhang, W.~D.~Guo and Y.~X.~Liu,
``Gravitational resonances in mimetic thick branes,''
JHEP \textbf{04} (2019), 154
[arXiv:1812.06453 [gr-qc]].

\bibitem{Bazeia:2012br}
D.~Bazeia, F.~A.~Brito and F.~G.~Costa,
``Braneworld solutions from scalar field in bimetric theory,''
Phys. Rev. D \textbf{87} (2013) no.6, 065007
[arXiv:1210.6318 [hep-th]].

\bibitem{Andrade:2018afh}
I.~Andrade, M.~A.~Marques and R.~Menezes,
``Cuscuton kinks and branes,''
Nucl. Phys. B \textbf{942} (2019), 188-204
[arXiv:1806.01923 [hep-th]].

\bibitem{Veras:2017nke}
D.~F.~S.~Veras and C.~A.~S.~Almeida,
Phys. Rev. D \textbf{95} (2017) no.10, 104032
doi:10.1103/PhysRevD.95.104032
[arXiv:1702.06263 [gr-qc]].

\end{thebibliography}
\end{document}